\documentclass{JAC2001}

\usepackage{graphicx}

\newcommand{\EPEM}{\mbox{e$^+$e$^-$}}

\newcommand{\GE}{\mbox{$\gamma$e}}

\newcommand{\TEV}{\mbox{TeV}}
\newcommand{\GEV}{\mbox{GeV}}

\newcommand{\CM}{\mbox{cm}}

\newcommand{\MKM}{\mbox{$\mu$m}}

\newcommand{\MRAD}{\mbox{mrad}}

\newcommand{\PE}{\mbox{${\cal{P}}_e$}}
\newcommand{\PG}{\mbox{${\cal{P}}_{\gamma}$}}

\begin{document}
\title{\vspace*{-1.cm} \hspace*{11cm}  {\large\rm BUDKER-INP 2003-14}  \\[-2mm]
\hspace*{10.2cm}  {\large\rm LC-DET-2003-054} \\[2mm]
BEAM ENERGY MEASUREMENT AT LINEAR COLLIDERS \\ 
USING SPIN PRECESSION~\thanks{Talk at 26-th Advanced ICFA Beam Dynamic
  Workshop on Nanometre-Size Colliding Beams (Nanobeam2002), Lausanne,
  Switzerland, Sept 2-6, 2002.}}

\author{V.I. Telnov \\
{\it Institute of Nuclear Physics, 630090 Novosibirsk, Russia}}

\pagestyle{plain}
\maketitle

\begin{abstract}
  Linear collider designs foresee some bends of about 5-10 mrad. The
  spin precession angle of one TeV electrons on 10 mrad bend is $23.2$
  rad and it changes proportional to the energy. Measurement of the
  spin direction using  Compton scattering of laser light on electrons
  before and after the bend allows determining the beam energy with an
  accuracy about of $10^{-5}$. In this paper  the principle of the
  method, the procedure of the measurement  and possible errors are
  discussed.  Some  remarks about importance of plasma focusing effects in the
  method of beam energy measurement using Moller scattering are given.
\end{abstract}

\vspace{-0.3cm}
\section{INTRODUCTION}
 Linear colliders are machines for precision measurement of particle
properties, therefore good knowledge of the beam energy is of great
importance.  At storage rings the energy  is calibrated by the method
of the resonant depolarization~\cite{BINP}. Using this method at LEP
the mass of $Z$-boson has been measured with an accuracy of $2.3\times
10^{-5}$~\cite{LEPZ}.  Recently, at VEPP-4 in Novosibirsk, an accuracy of
$\Psi$-meson mass of $4\times 10^{-6}$ has been achieved~\cite{KEDR}.
At linear colliders (LC) this method does not work and some other
techniques should be used.  The required knowledge of the beam energy
for the t-quark mass measurement is of the order of $10^{-4}$, for the
WW-boson pair threshold measurement it is $3\times 10^{-5}$ and ultimate energy
resolution, down to $10^{-6}$, is needed for new improved Z-mass
measurement. In other words, the accuracy should be as good as
possible.

In the TESLA project~\cite{TESLATDR} three methods for beam energy
measurement are considered: magnetic spectrometer\cite{SLCspect},
Moller (Bhabha) scattering~\cite{LEPM} and radiative return to
Z-pole~\cite{RAD}.  In the first method the accuracy $\Delta E/E
\sim 10^{-4}$ is feasible, if a Beam Position Monitor (BMP) resolution
of 100 nm is achieved. In the Moller scattering method an overall
error on the energy measurement of a few $10^{-5}$ is
expected~\cite{LEPM,TESLATDR}. However, the resolution of this method
may be much worse due to plasma focusing  effects in the gas jet, see
Sect.~\ref{Moller}. In order to decrease these effects the gas target
should be thin enough which results in a long measuring time. 

In this paper a new method of the beam energy measurement is
considered based on the precession of the electron spin in big-bend
regions at linear colliders.  It is not a completely new idea, after
success of the resonant depolarization method people asked whether
spin precession can be used for beam energy measurement at a linear
collider. However, nobody has considered this option
seriously~\cite{Torrence} (see also remark in Sect.\ref{7}).
 
\section{Principle of the method}
This method works if two conditions are fulfilled:
\begin{itemize}

\item electrons (and (or) positrons) at LC have a high a degree of
polarization. If a second beam is unpolarized its energy 
can be found from the energy of the first beam using the acollinearity
angle in  elastic \EPEM\ scattering.   
 
\item  there is a big (a few to ten mrads) bending angle between the
  linac and interaction point (IP). Such bend is natural in  case of
  two interaction regions and in the scheme with the crab-crossing,
  otherwise the angle about 5 mrad can be intentionally added to
  a design.
\end{itemize}

During the bend the electron spin  precesses around a vertical
magnetic field. The spin angle in respect to the direction of motion
$\theta_s$ varies proportionally to the bending angle
$\theta_b$~\cite{landau}
\begin{equation}
\theta_s = \frac{\mu^{\prime}}{\mu_0}\gamma \theta_b \approx \frac{\alpha
  \gamma}{2\pi}\, \theta_b,
\end{equation}
where $\mu_0$ and $\mu^{\prime}$ are normal and anomalous electron
magnetic momenta, $\gamma= E/m_ec^2$, $\alpha =e^2/\hbar c \approx
1/137$. For $E_0 = 1$ TeV and $\theta_b \sim 10$ mrad the spin rotation
angle is 23.2 rad. The energy is found by measuring $\theta_s$ and
$\theta_b$.

The bending angle $\theta_b$ is measured using geodesics methods and
beam position monitors (BPM), $\theta_s$ can be measured using the
Compton polarimeter which is sensitive to the longitudinal electron
polarization, i.e. to the projection of the spin vector to the
direction of motion.   Assuming that the bending angle is measured
very precisely (with relative accuracy smaller than the required energy
resolution), the resulting accuracy of the energy  is
\begin{equation}
\frac{\Delta E}{E_0} = \frac{\Delta \theta_s}{\theta_s} =
\frac{2\pi \Delta \theta_s}{\alpha \gamma \theta_b} \sim
\frac{0.43}{E_0(\TEV)\, \theta_b(\MRAD)}\Delta \theta_s\,.
\label{DeltaE}
\end{equation}
Possible accuracy of $\theta_s$ is discussed later.

\begin{figure*}[!hbt]
\centering
\hspace*{-0.0cm}  \includegraphics[width=16cm,angle=0,]{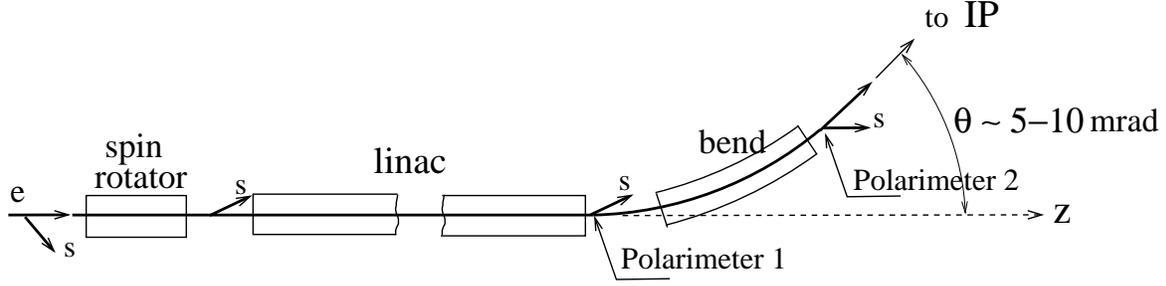}
\caption{Scheme of the energy measurement at  linear colliders using 
the spin precession.}
\label{escheme}
\end{figure*}

\begin{figure}[!hbt]
\centering
\hspace*{-0.0cm}  \includegraphics[width=8.5cm,angle=0,]{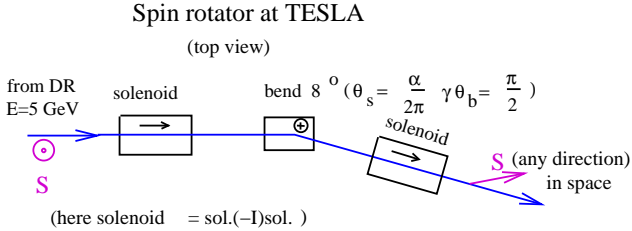}
\caption{Scheme of the spin rotator, top view.}
\label{rotator}
\end{figure}

A scheme of this method is shown in Fig.\ref{escheme}.
The spin rotator at the entrance to the main linac can make any spin
direction  conserving the absolute value of the polarization vector
{\boldmath $S$}. A scheme of the rotator in the TESLA project is shown
in Fig.\ref{rotator}. It consists of three sections: \\[-4mm]
\begin{itemize}
\item an initial solenoid unit, which rotates the spin around the local
  longitudinal (z)
  axis by $\pm 90 \, ^{\circ}$; \\[-4mm]
\item a horizontal arc which rotates the spin around the vertical
  axis by $90\,^{\circ}$ ($8^{\circ}$ bend for the 5 GeV beam energy
  after the damping ring);\\[-4mm]
\item a final solenoid unit providing an additional rotation about z-axis
  by $\pm 90\,^{\circ}$.\\[-4mm]
\end{itemize}
The solenoid unit consists of two identical solenoids separated by
short beamline whose (transverse) optics forms $(-I)$ transformation, thus
effectively cancels the betatron coupling while the spin rotation of
two solenoids add~\cite{TESLATDR}.

After the damping ring (DR) the electron spin {\boldmath $S$} has the vertical
direction (perpendicular to the page plane). At the exit of the spin
rotator it can have any direction.

In the considered  method the electron polarization vector  should be
oriented in the bending plane with  high accuracy. Two Compton
polarimeters measure the angle of the polarization vectors (before and
after the bend). This allows one to find the beam energy.

A Compton polarimeter was used at SLC~\cite{SLC} and other experiments
and will be used at the next LC for measurement of the longitudinal
beam polarization~\cite{TESLATDR}. The expected absolute accuracy of
polarimeters is $\leq {\cal{O}}(1\%)$, but the relative variation of the
polarization can be measured much more precisely.

\section{Measurement of the spin angle}

\begin{figure}[!hbt]
\centering
\hspace*{-0.0cm}  \includegraphics[width=8cm,angle=0,]{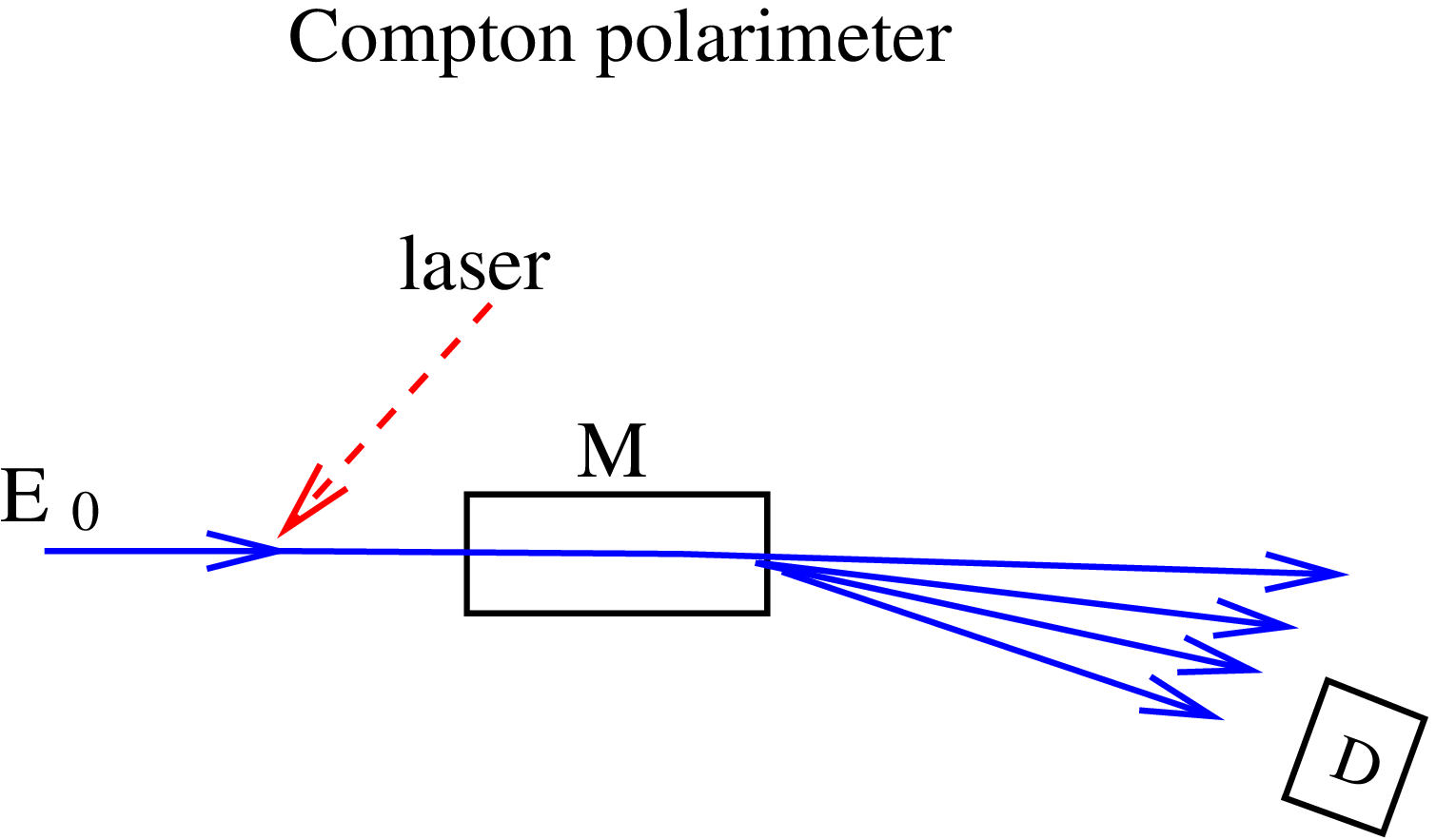}
\caption{Compton polarimeter. M is the analyzing magnet, D the detector
  of  electrons with large energy loss.}
\label{polarimeter}
\end{figure}

The longitudinal electron polarization is measured by Compton
scattering of circularly polarized laser photons on electrons. After
scattering off 1 eV laser photon the 500 GeV electron loses up to 90 \%
of its energy~\cite{GKST}, namely these low energy electrons are
detected for measurement of the polarization~(see
Fig.\ref{polarimeter})

The energy spectrum of the scattered electrons in collisions of
polarized electrons and photons is defined by the Compton cross
section~\cite{GKST84}
\begin{equation}
{d\sigma \over d y}  = {d\sigma_{u} \over d y}[1+\PG\PE\ F(y)],
 \;\;\;\;\;  y = \frac{E_0-E_e}{E_0},
\end{equation}
where $E_e$ is  the scattered electron energy, the unpolarized Compton
cross section
$$ {d\sigma_{u}\over d y}   = {2\sigma_{0}\over x}
\left[ {1\over 1-y} + 1-y - 4r(1-r) \right], $$
$$F(y)= \frac{rx(1-2r)(2-y)}{1/(1-y) +1-y-4r(1-r)},$$
$$x \approx \frac{4E_0\omega_0}{m^2c^4} = 19 \left[\frac{E_0}{\TEV}\right ]
\left[\frac{\MKM}{\lambda}\right],\;\;  r=\frac{y}{(1-y)x}\,,  $$
$$\sigma_{0}= \pi r_e^2 = \pi\left({e^2\over mc^2}\right)^2 = 2.5\times
10^{-25}\,\CM ^{2}\, ,$$ 
$\PE\ = 2\lambda_e  $ is the longitudinal electron polarization
(doubled mean electron helicity) and \PG\ is the photon helicity,
$\omega_0$ is  the laser photon energy, $\lambda$ the wavelength.
The minimum electron energy 
$E_{e,\,\min} = E_0/(x+1)$.

\begin{figure}[!hbt]
\vspace{-0.8cm}
\centering
\hspace*{-0.5cm}  \includegraphics[width=9cm,angle=0,]{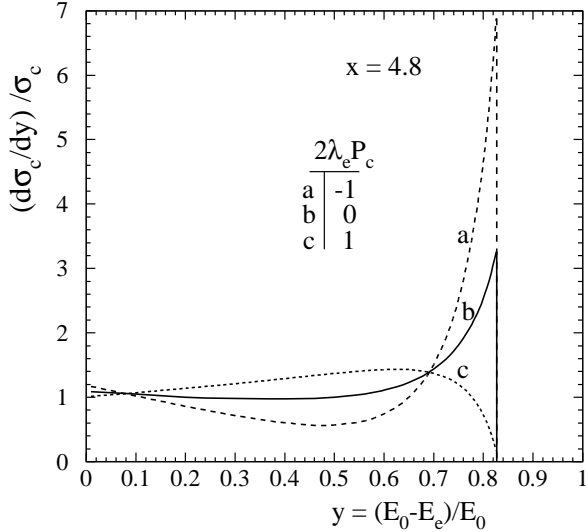}
\vspace{-1.5cm}
\caption{Spectrum of the Compton scattered electrons for various
  relative polarizations of laser end electron beams.}
\label{spectrum}
\end{figure}

For example, at $E_0  = 250$ GeV and $\lambda = 1\; \mu$m, $x \approx
4.8$,  the minimum electron energy is about $0.18E_0$. The scattered photon
spectra for this case are shown in Fig.\ref{spectrum}. 
If one detect the scattered electrons in the energy range close to the
minimum energies, the counting rate (or  the analog signal in the
polarimeter (see Sect.\ref{5.4}) which is better suited for our
task) is very sensitive to the product of laser and electron
helicities, see Fig.\ref{spectrum},
\begin{equation}
 \dot{N} \propto (1- \PG\PE) + {\cal{O}}(0.2-0.3), 
\label{dotN}
\end{equation}
here ${\cal{O}}$ means ``about''. In real experimental conditions  some
background is possible, according to estimates and previous experience
at SLC~\cite{SLC} it can be made small compared to the signal.

The longitudinal electron polarization is given by  $\PE\ = P_e \cos{\theta}$,
where $P_e$ is the absolute value of the polarization, $\theta$
the angle between the electron spin and  momentum. According to
(\ref{dotN}) the number of events in the polarimeter for a certain
time is
\begin{equation}
N = A \cos{\theta} + B\,,
\label{N}
\end{equation}
where $A \sim B$. This dependence is valid for all \GE\
processes~\cite{GKST84}, including Compton scattering with radiation
corrections.  Varying $\theta$ using the spin rotator one
can find $N_{\max}, N_{\min}$ corresponding to $\theta=0$ and $\pi$,
then for other spin directions the angle can be found from the
counting rate
\begin{equation}
\cos{\theta} = \frac{2N - (N_{\max}+N_{\min})}{N_{\max}-N_{\min}}\,.
\label{cos}
\end{equation}
Measurements of $\theta$ before ($\theta_1$) and after the bend
($\theta_2$) give the precession angle
\begin{equation}
\theta_s= \theta_{2} - \theta_{1}.
\label{ts}
\end{equation} 

\section{Statistical accuracy}

The statistical accuracy can be evaluated from (\ref{cos}). Assuming
that in  both polarimeters $|\sin{\theta}|$ are chosen to be large
enough (at any energy it is possible to make both $|\sin{\theta}| >
0.7$) and $N_{\min},N_{\max}$ and $N$ are measured, the statistical
accuracy of the precession angle is
\begin{equation}
\sigma(\theta_s) < \frac{5}{\sqrt{N}},
\end{equation}
where $N$ is the number of events in each polarimeter for the total
time of measurement. If the Compton scattering probability is
$10^{-7}$  and  30\% of scattered electrons with minimum
energies are detected, then the counting rate for TESLA is $2\cdot
10^{10} \times 14\,{\mbox{kHz}}\times 10^{-7} \times 0.3 = 10^7$ per
second. The statistical accuracy of $\theta_s$ for 10 minutes run is
$6 \times 10^{-5}$. To decrease systematic errors  one has to make
some additional measurements (see the next section),  which increase
the measuring time  roughly by factor of 3.  Using (\ref{DeltaE})
we can estimate the accuracy of the energy measurement for
1/2 hour run and $\theta_b = 10$ \MRAD\
 \begin{equation}
\frac{\Delta E}{E_0} \sim \frac{2.5\times 10^{-6} }{E_0[\TEV]}\,.
\end{equation}

It is not necessary to measure the energy all the time. During the
experiment one can make calibrations at several energies and then use
measurements of the magnetic fields in bending magnets  for
calculation of energies at intermediate energy points. Between the
calibrations it is necessary to check periodically the bending angle
and stability of magnetic fields in the bending magnets.

If one spends only 1\% of the  time for the
energy calibrations the overall statistical accuracy for $10^{7}$ sec
running time will be  much better  than $10^{-5}$ for any LC energy and
bending angles larger than several mrads.

In the experiment,  it is important also to know the energy of each
bunch in the train.  Certainly, the dependence of the energy on the
bunch number is smooth (after averaging over many trains)  and can be
fitted by some curve, therefore the energy of each bunch  will be
known  only somewhat worse than the average energy.
   
It seems that the statistical accuracy is not a limiting factor, the
accuracy will be determined by systematic errors.

\section{Procedure of the energy measurement \label{s5}}

Systematic errors depend essentially on the  procedure of  measurements. It
should account for the following  requirements:
\begin{itemize}
  
\item for the energy calibration  polarized electrons and circularly
  polarized laser photons are used, but  the result should not depend
  on the accuracy of the knowledge of their polarizations; 
\item  the measurement procedure includes some  spin manipulations
  using the spin rotator, the accuracy of
  such manipulation should not contribute to the result;
\item  change of the  spin rotator parameters may lead to some
  variations of the electron beam sizes,  position in the
  polarimeter and backgrounds, influence of these effects should be
  minimized.
   
\end{itemize}

Below we describe several procedures which can considerably reduce
possible systematic errors. 

\subsection{Measurement of $N_{\max},N_{\min}$  \label{s5.1}} 

The maximum and minimum signals in the polarimeter correspond to
$\theta=0$ or $\theta= \pi$, see (\ref{N}). To measure $N_{\max}$ one
can use the  knowledge of the accelerator properties and orient the
spin in the forward direction with some accuracy $\delta \theta$. Our
goal is to measure the  signal with an accuracy at the level of
$10^{-5}$. This needs $\cos{\,\delta\theta}<10^{-5}$ or  $\delta \theta
< 5\times 10^{-3}$. It is difficult to guarantee such accuracy,  it is
better to avoid this problem. The experimental procedure which allows
to reduce significantly this angle using minimum time is the
following.  In the first measurement instead of $\theta=0$ the spin
has some small unknown angles $\theta_x$ and $\theta_y$ in horizontal
and vertical planes, then the counting rate
\begin{equation}
N_{\max, 1} \approx A+B \cos{(\sqrt{\theta_x^2+\theta_y^2})} \approx A+B
(1-\theta_x^2/2 -\theta_y^2/2)\,.
\label{n1} 
\end{equation}
To exclude the uncertainty one can make some fixed {\it known}
variations  of $\theta_x$ and $\theta_y$ on about  $ 10^{-2}$ rads
based on knowledge of the spin rotator and  accelerator
parameters. The accuracy of such variations at the level of one
percent is more than sufficient. Eq.~(\ref{n1}) has 4 unknown
variables: $A,\, B,\, \theta_x,\, \theta_y$. To find them one needs 3
additional measurements. For example, in the second measurement one
can make the variation $\Delta \theta_x$, in the third minus $\Delta
\theta_x$ and in the fourth $\Delta \theta_y$. Solving the system of
four linear equations one can find  $\theta_x$, $\theta_y$, and after
that make the final correction using the spin rotator which places the
spin in the horizontal plane with very good accuracy (final angles are
about 100 times smaller than the initial $\theta_x$, $\theta_y$, if
the spin rotator makes the desired tilt with 1\% accuracy) and collect
larger statistics to determine $N_{\max}$. The minimum value of the
signal, $N_{\min}$, is found in a similar way making variations
around $\theta = \pi$.

\subsection{Positioning the spin to the bending plane}

For a precise measurement of the precession angle the spin should be
kept in the bending plane. Initially, one can put the spin in this
plane with an accuracy given by  the knowledge of the system. The
residual unknown angle $\theta_y$ can be excluded in a simple way. It
is clear that the {\it measured} precession angle is a symmetrical
function of $\theta_y$ and therefore depends on this small angle in a
parabolic way. Let us take three measurements of the precession angle
at  $\theta_y$ (unknown) and $\theta_y \pm \Delta \theta_y$.  These
three measurement give three values of the precession angle
$\theta_s(1)$, $\theta_s(2)$, $\theta_s(3)$ which correspond to three
equidistant values of $\theta_y$. After fitting the results by a
parabola one obtains the maximum  (or may be the minimum, depending on
the horizontal angles) value of $\theta_s$ which corresponds to  the
position of the spin vector in the bending plane. Using this result
one can place the spin to the bending plane using the spin rotator
with much higher accuracy and collect larger statistics for
measurement of $\theta_s$.

   Two additional remark to the later measurement:
\begin{enumerate}
\item The small vertical angle gives only the second order
  contribution to the precession angle   $\theta_s$, therefore the
  absolute values of the variations $\Delta\theta_y$ in the second and
  third measurements should be known with rather moderate accuracy.
  Furthermore, $\pm\Delta\theta_y$ give the scale and the final variation
  is taken as a certain part of $\Delta\theta_y$ (which is easier than
  some absolute value). For example, if $\Delta \theta_y \sim 3\times
  10^{-2}$ and  on the final step we add a
  part of this angle with an accuracy 3 \%, the final $\theta_y$ will be
  less than $10^{-3}$ ($< 5 \times 10^{-3}$ is needed, Sect.\ref{s5.1}). 
  
\item Varying $\theta_y$ one can make an uncontrolled variation of
  $\theta_x$ at the entrance to the bending system. However, it makes
  no problem since we measure {\it the difference} of the $\theta_x$
  measured before and after the bend.

\end{enumerate}
   
\subsection{Variation of electron beam sizes and position in polarimeters}

Geometrical parameters of the electron beam can depends
somewhat on  spin rotator  parameters.
 In existing designs of the spin rotators~\cite{TESLATDR}
these variations are compensated, but some residual effects can remain.
These dependences should be minimized by proper adjustment of the accelerator;
additionally they can be reduced by taking laser beam sizes much larger
than those of the electron beams.

 The laser-electron luminosity (proportional to 
Compton scattering probability)  is given by {\small
\begin{equation}
L = \frac{N_e N_L \nu}{4\pi\sqrt{(\sigma_{y,L}^2+\sigma_{y,e}^2)
[(\sigma_{z,L}^2+\sigma_{z,e}^2)(\theta/2)^2 + (\sigma_{x,L}^2+\sigma_{x,e}^2)]}},
\label{L1}
\end{equation}
} where $\theta$ is the collision angle, $\sigma_{i,e}$ are the
electron beam sizes, $\sigma_{i,L}$ are the laser beam sizes,
$N_e,\,N_L$ are the number of particles in the electron and laser
beams and $\nu$ is the beam collisions rate. This formula is valid when
the Rayleigh length $Z_R$ (the $\beta$-function of the laser beam) is
larger than the laser bunch length. Assuming that electron beam sizes
are much smaller than those of the laser, the  laser beam is round
($\sigma_{x,L}=\sigma_{y,L}$) and its sizes are  stable  we get { $$L=
  \frac{N_e N_{\gamma} \nu}{4\pi\sigma_{y,L} \sqrt{( \sigma_{z,L}^2
      (\theta/2)^2 + \sigma_{y,L}^2)}} \times$$
\begin{equation}
 \left(1-\frac{\sigma_{y,e}^2}{2\sigma_{y,L}^2} -
\frac{\sigma_{z,e}^2\theta^2 + 4\sigma_{x,e}^2}{2(\sigma_{z,L}^2\theta^2
  +  4\sigma_{y,L}^2)}\right) 
\label{L2}
\end{equation}}
Electron beam sizes at maximum LC energies (but not at the interaction
point) are of the order of  $\sigma_{z,e}=100-300$ \MKM, $\sigma_{x,e}
\sim 10\,\MKM$,\,  $\sigma_{y,e} \sim 1\,\MKM$. In order to
reduce the dependence on the electron beam parameters laser beam sizes
should be much larger than those of the electron beams, i.e.
$\sigma_{y,L} \gg \sigma_{y,e}$ and $\sigma_{z,L} \theta \gg
\sigma_{x,e}$. Under these conditions the collisions probability
depends on variations of the transverse electron beam sizes as follows
\begin{equation}
\frac{\Delta L}{L} = \left(\frac{\sigma_{y,e}}{\sigma_{y,L}}\right)^2 
\frac{\Delta \sigma_{y,e}}{\sigma_{y,e}} + 
 \left(\frac{2\sigma_{x,e}}{\sigma_{z,L}\, \theta}\right)^2 
\frac{\Delta \sigma_{x,e}}{\sigma_{x,e}}
\end{equation} 

Our goal is to measure the signal in the polarimeters with an accuracy
about $10^{-4}$. Let the transverse electron beam size varies on 10
\%. In order to decrease the corresponding variations of $L$  down to the
desired level one should take
\begin{equation}
\sigma_{y,L} = \sigma_{x,L} \approx 30 \,\sigma_{y,e} \sim 30\; \MKM,
\label{sLy}
\end{equation}
\begin{equation}
\sigma_{z,L} \theta  \approx 30\times  2\sigma_{x,e} \sim 600\,\MKM.
\label{sLz}
\end{equation}
 Deriving (\ref{L1}) we assumed $\sigma_{z,L} < Z_R$, the latter
  can be found from (\ref{sLy}) using the  relation $\sigma_{y,L} \equiv
  \sqrt{\lambda Z_R/4\pi}$. It gives
 \begin{equation} 
\sigma_{z,L} < Z_R = 4\pi\sigma^2_{y,L}/\lambda \sim 1\;\CM, 
\label{zL}
\end{equation}
where  $\lambda=1$ \MKM\ was  assumed. 

Eqs.(\ref{sLz}) and (\ref{zL}) do not fix the collision angle. As the
laser beam is cylindrical, the collision probability will be the same
if  one takes long bunch and small angle or short bunch and large
angle.  For example, in the considered case of $\sigma_{x,e} =
10\;\MKM$ and $\sigma_{y,e} = 1\;\MKM$,  one can take $\sigma_{z,L}
\sim 0.5 Z_R \sim 0.5$ cm (longest as possible according to (\ref{zL}))
and $\theta \sim 0.1$.

The required laser flash energy ($A$) can be found from
(\ref{L1}) and relations
$$
L\sigma_c = kN_ef \;\;\;\;\; A=\omega_0 N_{\gamma}\,,$$
where $k$
is the probability of Compton scattering (for electrons) and
$\sigma_c$ is the Compton cross section.  Leaving the
dominant laser terms which were assumed to be 30 times larger than the electron
beam sizes, we find the required laser flash energy
\begin{equation}
A \approx \omega_0 \frac{4\pi\sigma_{x,e}\sigma_{y,e} (30)^2 k}{\sigma_c}. 
\end{equation} 
For example, for $\lambda = 1\;\MKM$ ($\omega_0 = 1.24$ eV),
$\sigma_{x,e} = 10$ \MKM, $\sigma_{y,e} = 1$ \MKM, $k = 10^{-7}$ and
$\sigma_c = 1.7\times 10^{-25}\; \CM^2$ (for $E_0=250\;\GEV)$ we get
$A=1.3\times 10^{-4}$ J. The average laser power at 20 kHz collision
rate is 2.5 W (no problem).

   Another way to overcome this problem is a direct measurement of
   this effect and its further  correction. In this case the laser
   beam can be focused more tightly. In order to do this one should
   take the photon helicity  be equal to zero and change the electron
   spin orientation in the bending plane using the spin rotator. As
   the Compton cross section depends on the product of laser and
   electron circular polarization the signal in the polarimeters may be
   changed only due to the electron beam size effect.  To make sure
   that circular polarization of the laser {\it in the collisions
     point} is  zero with a very high accuracy one can take the
   electron beam with longitudinal polarization close to maximum and
   vary the helicity of laser photons using a Pockels cell. The
   helicity is zero when counting rate in the polarimeter is
   $0.5(N_{\max}+N_{\min})$.   These data can be used for correction
   of the residual  beam-size effect.
   
   The position of the electron beam in the polarimeters can be
   measured using beam position monitors (BPM) with a high accuracy.
   The trajectory can be kept stable for any spin rotator parameters using
   the BPM signals and corrector magnets.

\subsection{Detector \label{5.4}}

As a detector of the Compton scattered electrons one can use the  gas
Cherenkov detector successfully performed in the Compton polarimeter
at SLC~\cite{SLC}. It detects only particles traveling in the forward
direction and is blind for  wide angle background. The expected number
of particles in the detector from  one electron bunch is about
1000. Cherenkov light is detected by several photomultipliers.

To correct nonlinearities in the detector  one can use several
calibration light sources which can work in any combination  covering
the whole dynamic range.
   
   For accurate subtraction of variable backgrounds (constant
   background is not a problem) one can use events without
   laser flashes. Main source of background is bremsstrahlung on
   the gas. Its rate is smaller than from Compton scattering and does
   not present a problem.

\subsection{Measurement of the bending angle}

We assumed that the bending angle can be measured with negligibly
small accuracy. Indeed, beam position monitors can measure the
electron beam position with submicron accuracy. In this way one can
measure the direction of motion. Measurements of the angle between two
lines separated by several hundreds meters in air is not a simple
problem,  but there is no fundamental physics limitation at this
level. For example, gyroscopes (with correction to Earth
rotation) provide the needed accuracy.

\section{Systematic errors}
   
Some possible sources of systematic errors were discussed in the previous
section. Realistic estimation  can be done only after the
experiment. Measurement of $\Delta \theta_s$ (averaged over many pulses) on the
level $10^{-4}$ does not look unrealistic. The statistical accuracy
can be several times better and allows to see some possible systematic
errors.
  
If systematics are on the level $10^{-4}$, the accuracy of the energy
calibration according to (\ref{DeltaE}) is about
\begin{equation}
\frac{\sigma_E}{E_0} \sim  \frac{0.5\times 10^{-4}}{\theta_b[\MRAD] E_0 [\TEV]}.
\end{equation}

\section{Measurement of the magnetic field vs spin precession.} \label{7}

  There is a good question to be asked: maybe it is easier to measure
  magnetic field in all bending magnets instead of measurement of the
  spin  precession angle~\cite{Torrence}?
  
  Yes, it is more a straightforward way. However, we discuss the
  method which potentially allows an accuracy of the LC energy
  measurement of about $10^{-5}$. Bending magnets in the
  big-bends should be weak enough, $B \sim 10^3$ G, to preserve small
  energy spread and emittances. Who can guarantee $10^{-2}$ G accuracy
  of the magnetic field when the Earth field is about 1 G?

\section{Some remarks on the beam energy measurement using Moller
  (Bhabha) scattering} 
\label{Moller} 
In this method electrons are scattered on electrons of a gas target,
the energy is measured using angles and energies of both final
electrons in a small angle detector~\cite{LEPM,TESLATDR}. For LEP-2
energy the estimated precision was about 2 MeV.

Here I would like to pay attention to one effect in this method which
was not discussed yet. It is a plasma focusing of electrons. The
electron beam ionizes the gas target, free electron quickly leave the
beam volume while ions begin to focus electrons. Deflection of
electrons in the ion field can destroy the beam quality and affect the energy
resolution.

Let us make some  estimations of this effect for $E_0=90$ GeV which
was considered in the original proposal for LEP-2~\cite{LEPM}, but for
linear collider beams. The angle of the scattered electron for the
symmetric scattering is $\theta = \sqrt{2m_e c^2/E} \sim 3 $ mrad. The
Moller (Bhabha) cross section for the forward detector  considered in
~\cite{LEPM}  is 15 (4) $\mu$b. The dominant contribution to the
energy spread of {\it measured} energy is due to the Fermi motion of the target
electron~\cite{LEPM}:  $\sigma_E/E = 3.6\cdot 10^{-3}$. Somewhat
smaller contribution gives the intrinsic beam energy spread. Let us
take the combined energy resolution (for one event) to be equal to
$\sigma_E/E = 5\cdot 10^{-3}$. In order to obtain 0.5 MeV statistical
accuracy in $10^3$ sec the luminosity of beam interactions with the
$H_2$ target should be about $L=0.6\,(2.4)\cdot 10^{32}$
cm$^{-2}$s$^{-1}$ for $e^-$ ($e^+$), or approximately $10^{32}$ (in
\cite{LEPM} $L=4\cdot 10^{31}$ was assumed). 

The luminosity is $L=N_e\nu n l$, where $N_e \sim 10^{10}$ is the
number of particles in the electron bunch, $\nu \sim 10^4$  the
collision rate, $n$ the density of electrons in the target and $l$ is
the target thickness. This gives the required depth of the gas target
$n\,l \sim 10^{18}$ cm$^{-2}$.

   Let us consider now ionization of the hydrogen target by the electron
   beam. The relativistic particle produces in $H_2$ at normal pressure
   about 8.3 ions/cm, this corresponds to the cross section (per one
   electron) $\sigma_i = 8.3/(2\times 2.68\cdot 10^{19})= 1.5\cdot
   10^{-19} \;$cm$^2$. The total number of ions produced by the beam
   $N_i = N_e \sigma_i n l = 10^{10} \times 1.5\cdot 10^{-19} \times
   10^{18} = 1.5\cdot 10^{9}$, that is 15\% of the number of particles
   in the beam.   
   
   For the vertical (smallest) transverse beam size smaller than the plasma
   wavelength  and the density of the beam higher than the plasma
   density, all plasma electrons are pushed out from the beam. These
   conditions correspond to our case. The maximum deflection angle of the
   beam electrons in the ion field is
\begin{equation}
\Delta\theta \sim \frac{2 r_e N_i}{\sigma_x \gamma}.
\label{defl}
\end{equation} 
The horizontal beam size $\sigma_x=\sqrt{\epsilon_{n,x} \beta/\gamma}
\sim \sqrt{3\cdot 10^{-4} \times 3000 / 2\cdot 10^{5}} \sim 2\times
10^{-3}$ cm. Here we assumed that $\beta \sim
300\,\sqrt{E_0\mbox{(GeV)}}$ cm. The resulting deflection angle is $\Delta
\theta \sim 2\cdot 10^{-6}$.
 
The energy resolution (systematic error) due to the plasma focusing is
$\sigma_E/E \sim 2 \Delta\theta/\theta \sim 1.4\times 10^{-3}$, that
two order of magnitude larger than our  goal (about $10^{-5}$).

  The angular spread in the beam in the vertical direction is 
$\sqrt{\epsilon_{n,y}/(\beta_y \gamma)} \sim  \sqrt{3\cdot
  10^{-6}/(3000 \times 2\cdot 10^5)} \sim 0.7\cdot 10^{-7}$ rad that
is 30 times smaller than the deflection angle, so the beam after the
gas jet can not be used for the experiment.  

To avoid these problems one can take the gas target thinner by two
orders of magnitude. Then in the considered example the statistical
accuracy $10^{-5}$ for electrons is achieved in 4.5 hours.  Note that
at such beam thickness one can measure the energy and run experiment
simultaneously.

The cross sections of the Moller and Bhabha scattering depends on the
energy as $1/E^2$ which leads to increase of the measuring time for
higher energy. However one can increase the target thickness and allow
some degradation of the resolution. The optimum is reached when the
statistical error is equal to the systematic one. The systematic error
is $\Delta E/E \propto \Delta \theta/\theta \propto  nl/(\sigma_x
\gamma) /(1/\sqrt{\gamma}) \propto nl/\gamma^{1/4}.$ The statistical
error is $\sigma_E/E \propto 1/\sqrt{N} \propto 1/\sqrt{n\sigma l\,t}
\propto \gamma/\sqrt{nlt}$. At optimum conditions $\sigma_E/E \propto
\gamma^{7/12}/t^{1/3}$. So, for the same scanning time, a ten times
increase of energy leads to 3.8 times increase of the energy resolution.

  Several additional remarks on plasma effects which were not
  discussed here but may be important: 
\begin{itemize}
\item For positrons plasma effects are smaller because the ionization is
  confined in the beam channel and the scattered electrons after a short
  travel in the gas target get $\Delta y>\sigma_y$ where the ion and
  electron fields cancel each other;
\item in the above consideration the secondary ionization in the beam
  field was ignored; 
\item it is well known that short beams in plasma create strong
  longitudinal wakefields, about $E_z \sim \sqrt{n_p\,\mbox{[cm$^{-3}$]}}$
  eV/cm, which decelerates the beam. This effect may be not negligible
  in the considered problem.
\end{itemize}

\section{Discussion and Conclusion}

The method of beam energy measurement at linear colliders using spin
precession has been considered. The accuracy on the level of a few $10^{-5}$
looks possible. 

In this paper we considered only the measurement of the {\it average}
beam energy {\it before} the beam collision. Experiments will require
not this energy but the distribution of collisions on the invariant
mass. The beam energy spread at linear colliders is typically about
$10^{-3}$, but much larger energy spread and the shift of the energy
gives beamstrahlung during the beam collision. An additional spread in
the invariant mass distribution gives also an initial state radiation. So,
the luminosity spectrum will consist of the narrow peak with the width
determined by the initial beam energy spread and the tail due to
beamstrahlung and initial state radiation. This spectrum in {\it
  relative} units can be measured from the acollinearity of Bhabha
events~\cite{Miller1,Monig,Miller2}. The {\it absolute} energy scale is found
from the measurement of average beam energy {\it before} the beam
collisions which was discussed in the present paper. Namely the narrow peak
in the luminosity spectrum provides such correspondence. The
statistical accuracy of the acollinearity angle technique is high,
some questions  remain about systematic effects.

\section*{Acknowledgements}

I would like to thank Karsten Buesser and Frank Zimmermann for reading
the manuscript and useful remarks. This work was supported in part 
by INTAS 00-00679.

\end{document}